\documentclass{article}

\pdfoutput=1

\usepackage[final,nonatbib]{neurips_2020}
\usepackage{multirow}
\usepackage[utf8]{inputenc} 
\usepackage[T1]{fontenc}    
\usepackage{hyperref}       
\usepackage{url}            
\usepackage{booktabs}       
\usepackage{amsfonts}       
\usepackage{nicefrac}       
\usepackage{microtype}      
\usepackage{caption} 
\usepackage{lipsum} 
\usepackage{graphicx}
\usepackage{subcaption}
\usepackage{float}
\usepackage{wrapfig}
\usepackage[title]{appendix}

\title{Interpretable Poverty Mapping using Social Media Data, Satellite Images, and Geospatial Information}

\author{%
  Chiara Ledesma, Oshean Lee Garonita, Lorenzo Jaime Flores, Isabelle Tingzon, \\ \textbf{Danielle Dalisay}  \\
  Thinking Machines Data Science\\
  \texttt{$\{$chiara, oshean, lorenzo, issa, dani$\}$@thinkingmachin.es} \\
}

\begin{document}

\maketitle


\begin{abstract}
Access to accurate, granular, and up-to-date poverty data is essential for humanitarian organizations to identify vulnerable areas for poverty alleviation efforts. Recent works have shown success in combining computer vision and satellite imagery for poverty estimation; however, the cost of acquiring high-resolution images coupled with black box models can be a barrier to adoption for many development organizations. In this study, we present a interpretable and cost-efficient approach to poverty estimation using machine learning and readily accessible data sources including social media data, low-resolution satellite images, and volunteered geographic information. Using our method, we achieve an $R^2$ of 0.66 for wealth estimation in the Philippines, 
compared to 0.63 using satellite imagery. Finally, we use feature importance analysis to identify the highest contributing features both globally and locally to help decision makers gain deeper insights into poverty.

\end{abstract}

\section{Introduction}

The Zero Extreme Poverty Philippines 2030 (ZEP) movement aims to reduce poverty and improve the lives of millions of Filipinos by 2030 \cite{zep2030}. While partner organizations like UNDP have taken steps towards developing inclusive and sustainable programs, extreme poverty remains a major challenge in the Philippines. In 2018, nearly 17$\%$ of Filipinos were estimated to have lived below the national poverty line – a number that may increase further with the economic impacts of the COVID-19 pandemic \cite{adb2018}. With rampant poverty and inequality, humanitarian organizations face the challenge of having to quickly and efficiently identify vulnerable populations in order to strategically plan interventions for poverty reduction. 

Traditionally, data collection on poverty statistics is done through on-the-ground household surveys. Unfortunately, the expensive labor and long time frames required to conduct extensive surveys make it difficult for many developing countries to obtain complete and up-to-date socioeconomic data \cite{engstrom2016poverty,jerven2014benefits}. Several research works have sought to address this problem by using a combination of computer vision and satellite images to provide reliable and granular poverty estimates \cite{xie2016transfer,jean2016combining,tingzon2019mapping,ijcai2020-608,bai2020siamese,han2020learning,ni2020investigation}. However, the high costs associated with acquiring high-resolution satellite images combined with compute-intensive deep learning methods \cite{jean2016combining,tingzon2019mapping,ijcai2020-608,bai2020siamese} can be a barrier to adoption for many development organizations with limited resources. Moreover, the black box nature of deep learning models makes it difficult for decision makers and policymakers to understand model behavior and interpret results. Model interpretability is critical in AI-assisted decision-making processes for the development of policies and interventions that would significantly impact people's quality of life \cite{carvalho2019machine, rudin2019}.

In this study, we overcome these challenges by combining social media and geospatial data sources with cost-efficient machine learning methods as an interpretable and inexpensive approach to poverty estimation. Specifically, we examine the extent to which interpretable features derived from social media advertising data, remote sensing data, and volunteered geographic data can be used to estimate socioeconomic well-being in the Philippines. We demonstrate that not only does our approach outperform previous performance benchmarks for poverty estimation in the Philippines, but it also yields intuitive model explanations that can easily be understood by non-technical decision makers. 

\section{Datasets}

\subsection{Ground truth data}
In our work, we used the 2017 Philippine Demographic and Health Survey (DHS) which contains 27,496 households that are grouped into 1,249 household clusters \cite{dhs2017}. Cluster centroids are calculated as the average latitudes and longitudes of the households per cluster, with each centroid offset by 2 to 5 kilometers for data privacy protection. In line with the Zero Extreme Poverty themes \cite{zep2030}, we focused our analysis on a subset of survey questions related to asset ownership, water availability and access, toilet facility access, and educational attainment.

To align with previous studies \cite{tingzon2019mapping, fatehkia2020mapping}, we take the \textit{wealth index}, calculated as the first principal component of a number of asset-based variables, as our primary measure of socioeconomic well-being. Other derived target indicators include: proportion of households with toilet facilities located outside (toilet access); proportion of households with improved drinking water source (clean water source); proportion of household heads with higher educational attainment (educational attainment). 

\subsection{Geospatial covariates}
We derived a rich set of intuitive features from a combination of multiple, disparate datasets composed of social media data, remote sensing data, and point of interest (POI) data. For each dataset, we obtain the relevant features that lie within a 2 kilometer (urban) or 5 kilometer (rural) radius of each DHS cluster centroid. We discuss these datasets in more detail as follows:

\begin{itemize}
    \item \textbf{Social media advertising data.} Pew Research Center surveys report a relatively high Facebook penetration rate in the Philippines, with approximately 58$\%$ of adults using the social media platform \cite{pew2019}. Extracted data from Facebook’s Marketing API contains an approximate count of Facebook users within a specified radius from a set of coordinates. We used this to obtain the approximate number of Facebook users per DHS household cluster, with a breakdown of user segments such as users with 4G access, 3G access, 2G access, WiFi access, Apple devices, and mid-to-high valued goods consumer preferences.
    \item \textbf{Remote sensing data.} Using Google Earth Engine, we extracted features from publicly accessible low-resolution satellite images including: (1) nighttime luminosity data taken from the Visible Infrared Imaging Radiometer Suite provided by NASA, (2) daytime and nighttime land surface temperature derived from MODIS Satellite 2017 data, and (3) Normalized Difference Vegetation Index (NDVI) derived from Landsat 2017. For each satellite image, we computed summary statistics, i.e. the mean, maximum, minimum, skewness, variance, and kurtosis of all cloudless pixel values within each DHS cluster. 
    \item \textbf{Point of interest data.} Using OpenStreetMap (OSM), we obtained volunteered geographic information related to the counts of various points of interest, e.g. banks, restaurants, convenience stores, within each DHS household cluster. We also accessed hospital data from the Philippines' Department of Health as well as public school information from CheckMySchool, an education monitoring initiative in the Philippines \cite{nhfr, checkmyschool}. 
\end{itemize}

\section{Models}
\label{methods}
We evaluate the performance of the following regression models: Linear Regression, Lasso Regression, Ridge Regression, Random Forest, and LightGBM. Specifically, we trained the models on social media data, remote sensing data, and point of interest data, first separately then combined, with the hypothesis that integrating multiple data sources will lead to improved model performance over using any one data source alone. To retain the most important features, we used recursive feature elimination for feature selection and tuned hyperparameters using random search. All models were evaluated using a 5-fold cross validation approach.


\section{Results and discussion}

\begin{figure}[h]
    \vspace{-30pt}
    \centering
    \begin{subfigure}{0.35\textwidth}
    \includegraphics[width=\textwidth]{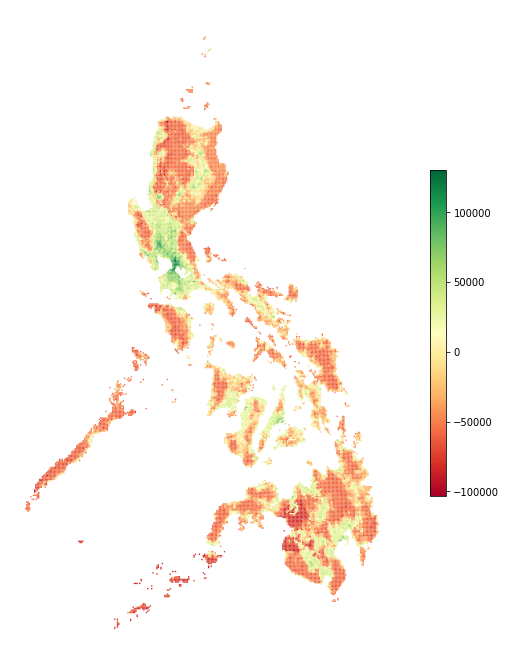}
    \caption{}
    \end{subfigure}
    \begin{subfigure}{0.41\textwidth}
    \includegraphics[width=\textwidth]{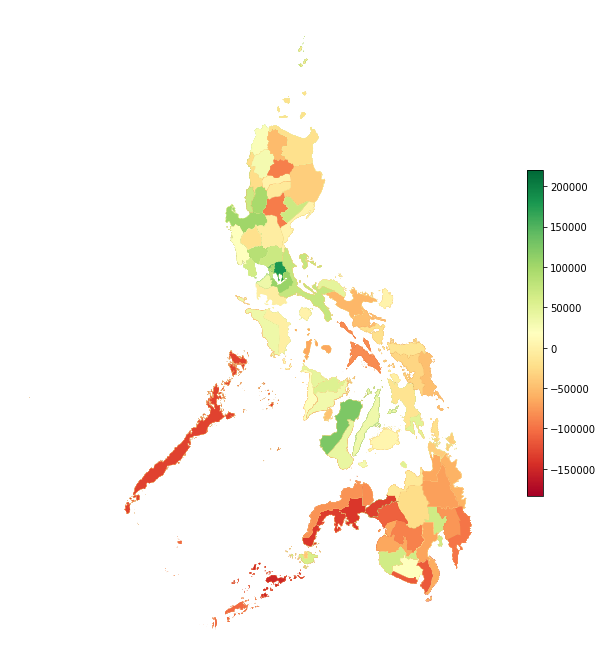}
    \caption{}
    \end{subfigure}
    \caption{(a) High resolution map of the 2017 Philippines DHS estimated wealth indices, and (b) province-aggregated 2017 Philippines DHS ground-truth wealth indices.}
    \label{fig:ph}
\end{figure}

In Table \ref{ml-results}, we compare the $R^2$ results of the different models described in Section \ref{methods} and find that the random forest regression model performs best for wealth estimation, with an $R^2$ of 0.66. Given the nature of the wealth variable as a constructed index, we chose $R^2$ as the most appropriate metric for interpretability.  A side-by-side comparison of the resulting high resolution wealth estimates and province-aggregated ground truth wealth indices are presented in Figure \ref{fig:ph}. Our findings indicate that combining the strengths of satellite data's truly global coverage with Facebook's asset ownership data as well as extensive volunteered geographic information yields better predictive accuracy than using any one data source alone. We note that our best result for wealth estimation marginally exceeds past performance benchmarks for the same task \cite{fatehkia2020mapping, tingzon2019mapping}. 

\begin{table}[h]
    \caption{Model comparison and $R^2$ for wealth estimation using social media data only (SM), remote sensing data only (RS), point of interest data only (POI), and a combination of all three (All).}
    \label{results-table}
    \centering
\begin{tabular}{lcccc}
\toprule
\multicolumn{1}{l}{\multirow{2}{*}{\textbf{Models}}} & \multicolumn{4}{c}{\textbf{$R^2$ scores}} 
\\ 
\cline{2-5} 
\multicolumn{1}{c}{}                                & \multicolumn{1}{l}{\textbf{SM}} & \multicolumn{1}{l}{\textbf{RS}} & \multicolumn{1}{l}{\textbf{POI}} & \textbf{All}                   \\ \midrule
Linear Regression        & 0.54                                & 0.44                                & 0.00                                 & 0.57                   \\
Lasso Regression         & 0.54                                & 0.44                                & 0.19                                 & 0.56                   \\
Ridge Regression         & 0.54                                & 0.44                                & 0.20                                 & 0.55                   \\
LightGBM Regression      & 0.55                                & 0.56                               & 0.49                                 & 0.64                   \\
\textbf{Random Forest Regression} & 0.55                               & 0.59                                & 0.49                                 & \textbf{0.66}                   \\ \midrule
Fatehkia, et al. \cite{fatehkia2020mapping} &        -                        &                          -       &    -                              & 0.63         \\
Tingzon, et al. \cite{tingzon2019mapping} &            -                    &                -                 &    -                              & 0.63         \\ \bottomrule
\end{tabular}
\label{ml-results}
\end{table}

Finally, we present in Figure \ref{fig:results} the cross validated predictions for all other development indicators estimated in the study. Consistent with previous works, we find that our approach does not generalize to other target indicators with the same level of accuracy as the wealth index. Specifically, we attained an $R^2$ of 0.58 for toilet access, 0.37 for access to clean water, and 0.35 for educational attainment. 

\subsection{Model interpretation}
\begin{figure}[h]
    \centering
    \includegraphics[width=0.57\textwidth]{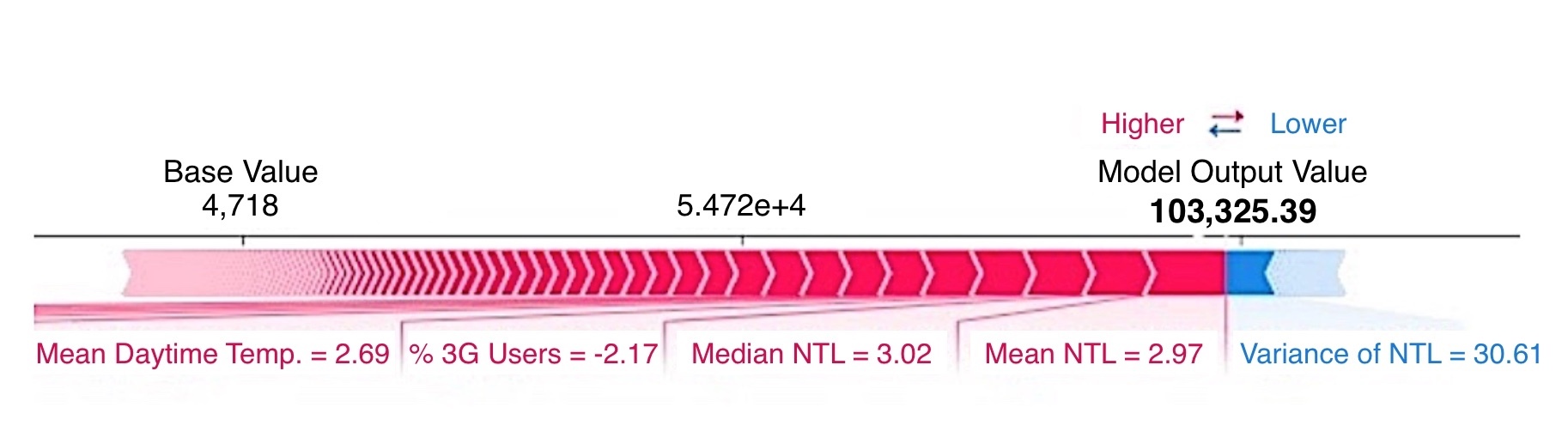}
    \caption{Example of SHAP values for predicted wealth index of one DHS cluster. The length of each arrow represents a variable’s contribution to a prediction, while the color represents whether a variable increased the predicted value (red) or decreased the value (blue).}
    \label{fig:s1}
\end{figure}

For model explainability, we used Shapley Additive Explanations (SHAP) to quantify the impact of each feature on a single prediction. The SHAP value for a variable $x$ is computed as the average difference between the model predictions trained with and without $x$ over all subsets of variables \cite{NIPS2017_7062, lundberg2018}. As an example, we show in Figure \ref{fig:s1} the SHAP values for the predicted wealth index of one DHS cluster in Marikina City. Here, the mean nighttime light values and percentage of 3G users both increase the predicted wealth index above the average score of 4,718. On the other hand, the variance in the nighttime light values within the geographic cluster leads to a slight decrease in the predicted wealth index.

\begin{wrapfigure}{R}{0.5\textwidth}
    \vspace{-20pt}
    \centering
    \includegraphics[width=0.5\textwidth]{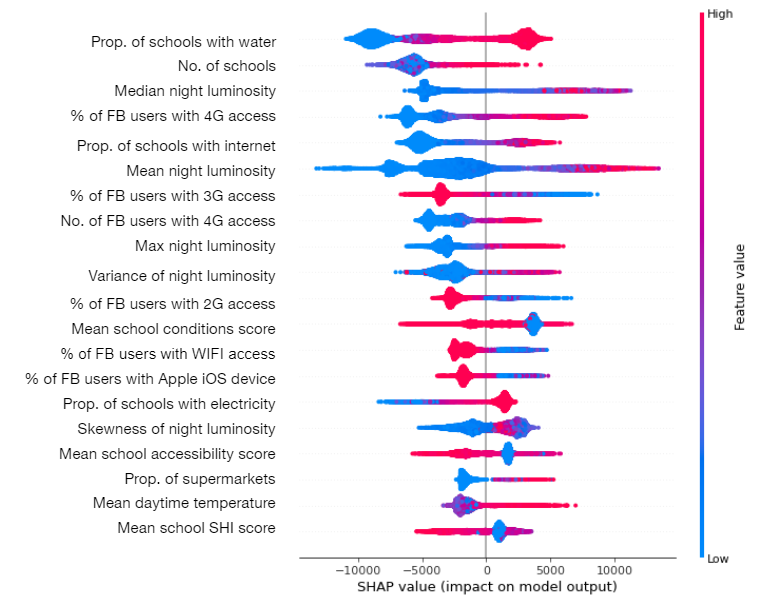}
    \caption{Summary of SHAP values for all clusters. Colors indicate relationship between variables and wealth index: blue to red signifies positive correlation with wealth, while red to blue signifies negative correlation.}
    \label{fig:s2}
\end{wrapfigure}

We also computed SHAP values for all observations, as shown in Figure \ref{fig:s2}. Variables with wider spread have a greater impact on each prediction, and we identified the variables with the greatest impact to be night time luminosity, proportion of 4G users, and the proportion of public schools with access to water. The results generally agree with domain knowledge of the on-the-ground situation in the Philippines and other poverty-related studies. Higher percentages of 4G users indicate greater wealth, as 4G access is significantly more expensive in the Philippines compared to low-quality internet connection such as 2G and 3G \cite{hasbi2020determinants, opensignal2019}. Wealthier areas such as urban cities also tend to have a higher concentration of establishments like supermarkets \cite{Kaufman:289786}. Finally, remote sensing data shows that wealthier areas tend to have brighter nighttime lights and higher temperature (urban heat island effect), as commonly observed in urban regions as opposed to rural areas \cite{Chen8589, li2013synergistic, yeh2020}.

\section{Conclusion}

In this study, we have proposed an alternative approach to poverty estimation using cost-efficient machine learning methods combined with social media data, remote sensing data, and point of interest data. Our approach improves over previous works in terms of model performance, cost of data acquisition, and model interpretability. Leveraging readily accessible datasets, we achieve an $R^2$ of 0.66 for wealth estimation using a simple random forest model. Finally, using SHAP for model interpretability, we determine the most predictive global indicators of wealth to be average night time light values, proportion of population with 4G access, and presence of public schools. Interpretable poverty maps are essential for policymakers and development organizations to understand the spatial distribution of poverty and to formulate targeted humanitarian programs and interventions. 

We note that a primary limitation of this study is reliance on incomplete and possibly unrepresentative data sourced from Facebook and OSM. While geospatial features derived from these datasets were found to correlate with wealth, we urge caution in interpreting these values and advise against making causal claims. We believe that further research is necessary to investigate the spatial coverage and completeness of these datasets. Future studies should also explore the applicability of these methods in other developing countries with potentially lower data completeness.  

\paragraph{Acknowledgements}
This work was made possible by funding from the United Nations Development Programme (UNDP)  through the Accelerator Lab Philippines and the Localizing e-Government for Accelerated Provision of the Services (LEAPS) project. We also thank our collaborators in Zero Extreme Poverty 2030 Philippines, who are using this research to pursue an evidence-based approach to lifting one million Filipino families from extreme poverty. 

\bibliography{bibliography} 
\bibliographystyle{plain}

\begin{appendices}
\section{Poverty Estimation Scatterplots}

\begin{figure*}[h]
  \centering
  \begin{subfigure}{0.24\textwidth}
  \includegraphics[width=\textwidth]{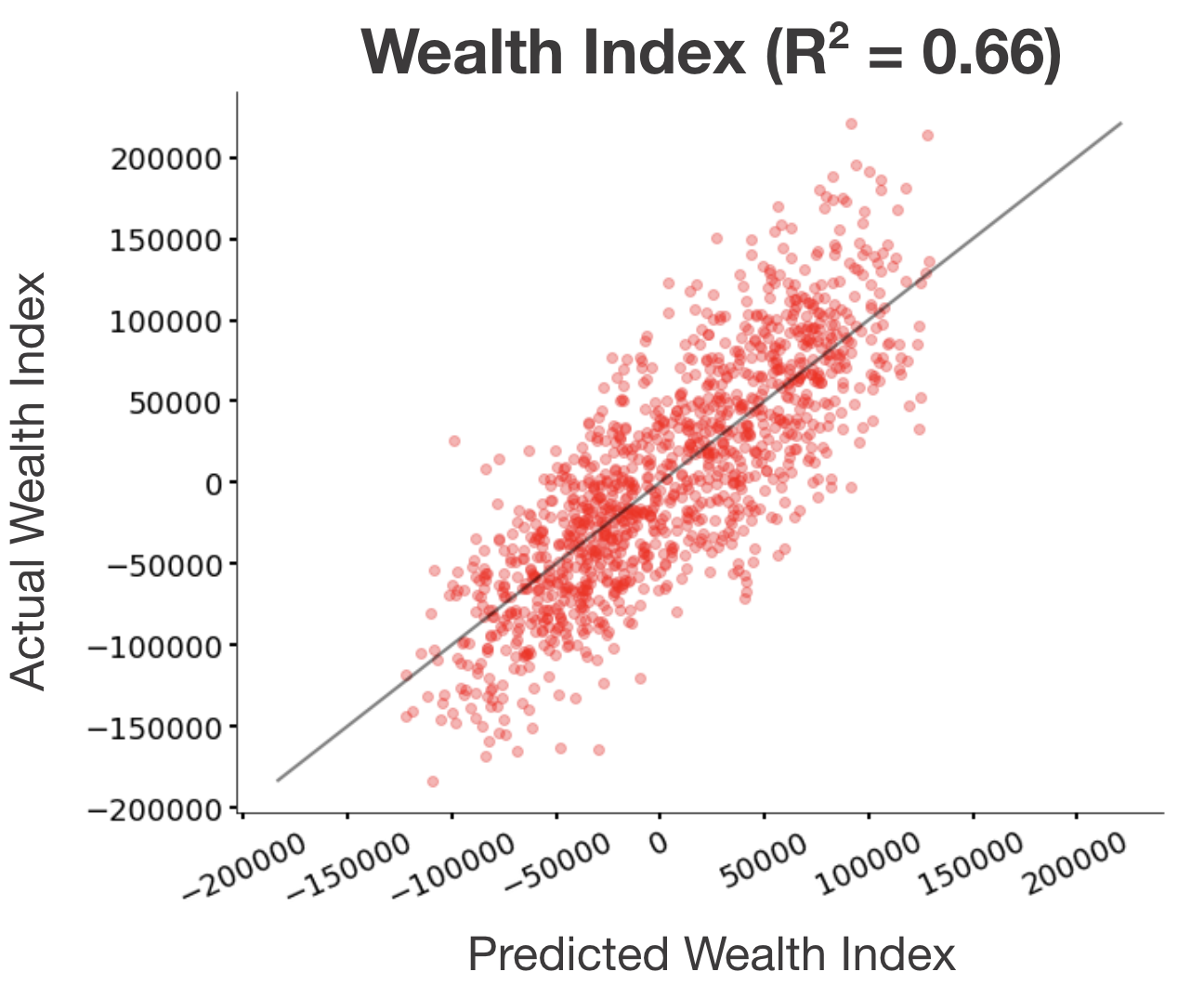}
  \end{subfigure}
  \begin{subfigure}{0.24\textwidth}
  \includegraphics[width=\textwidth]{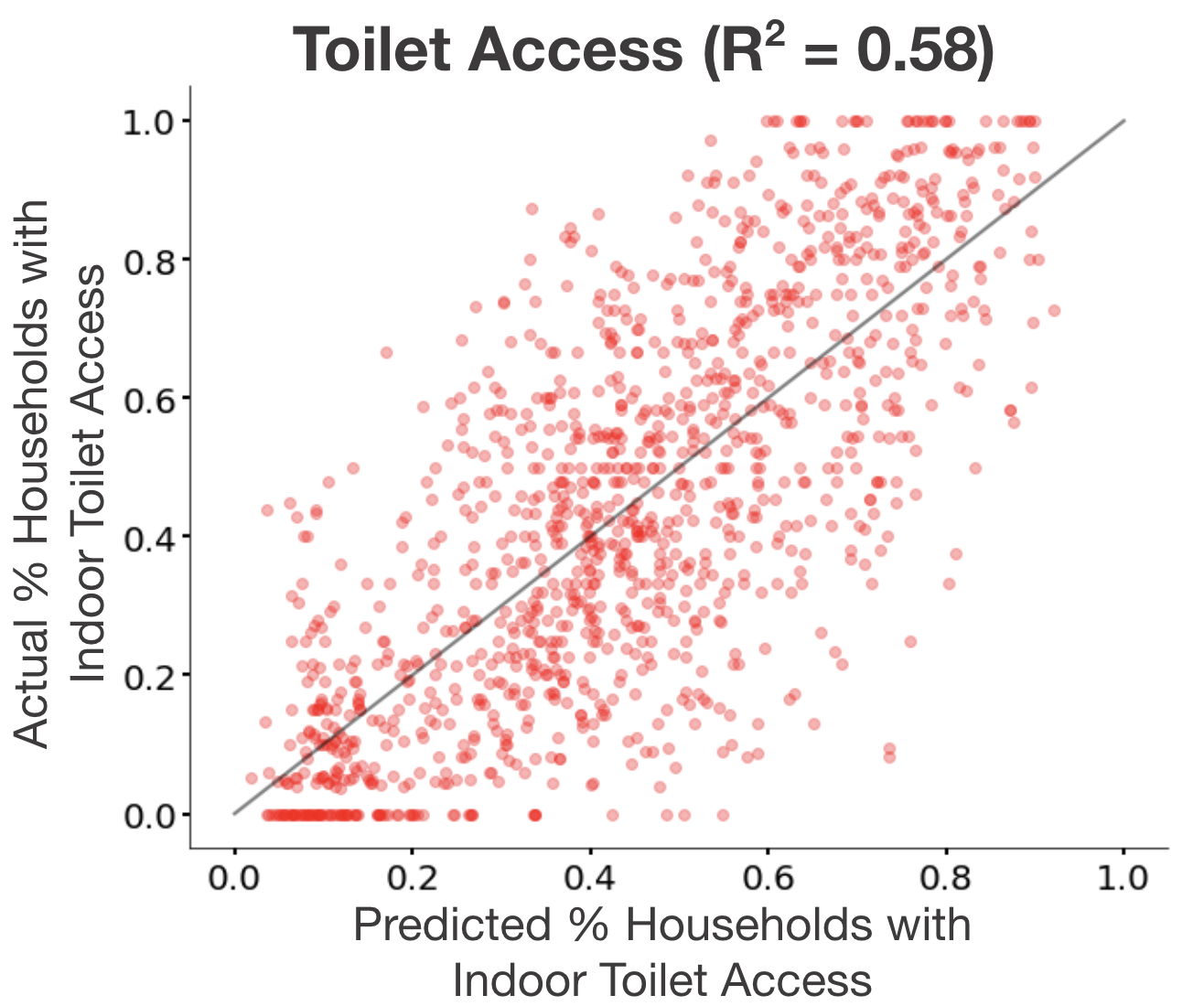}
  \end{subfigure}
  \begin{subfigure}{0.24\textwidth}
\includegraphics[width=\textwidth]{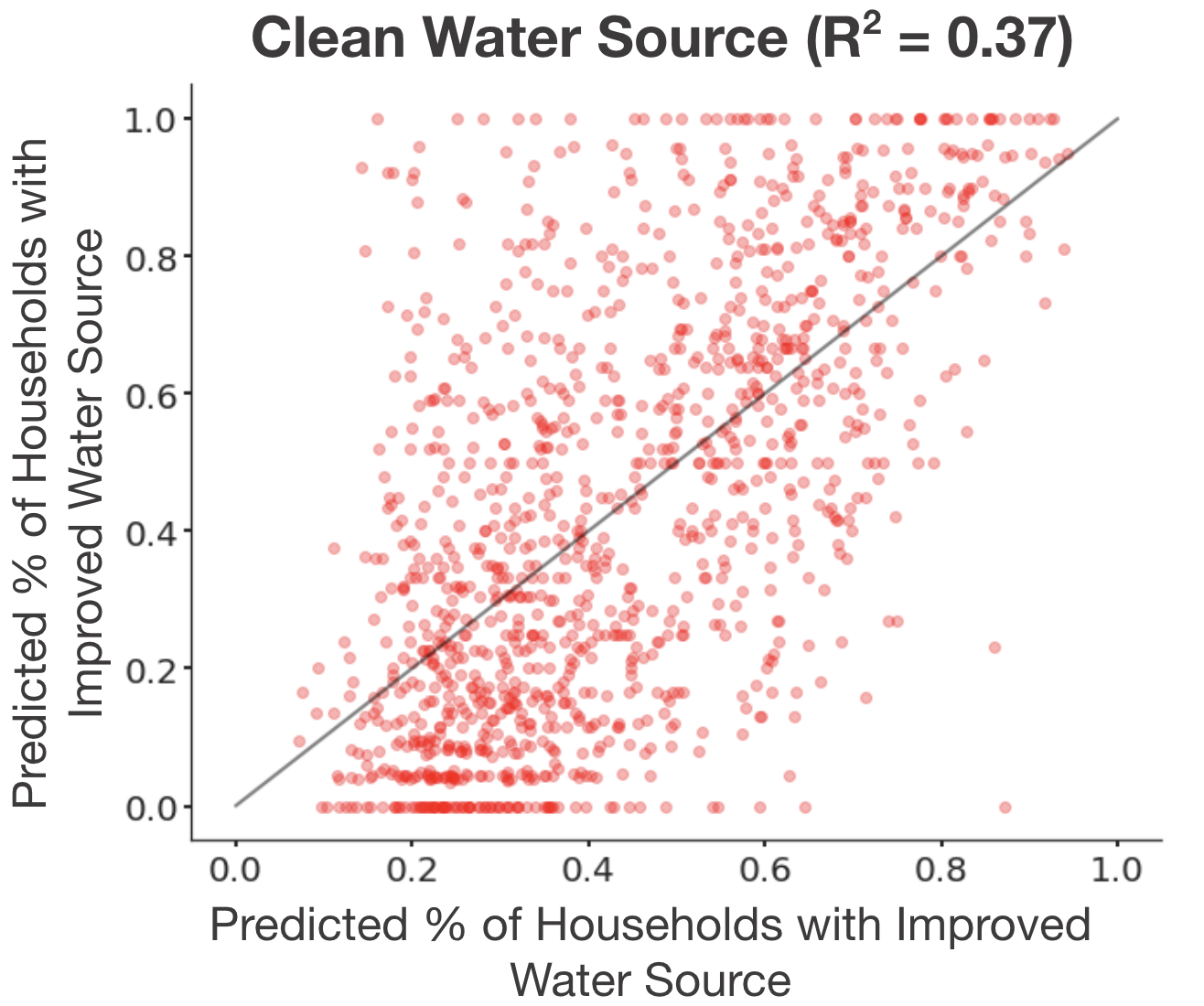}
  \end{subfigure}
  \begin{subfigure}{0.24\textwidth}
  \includegraphics[width=\textwidth]{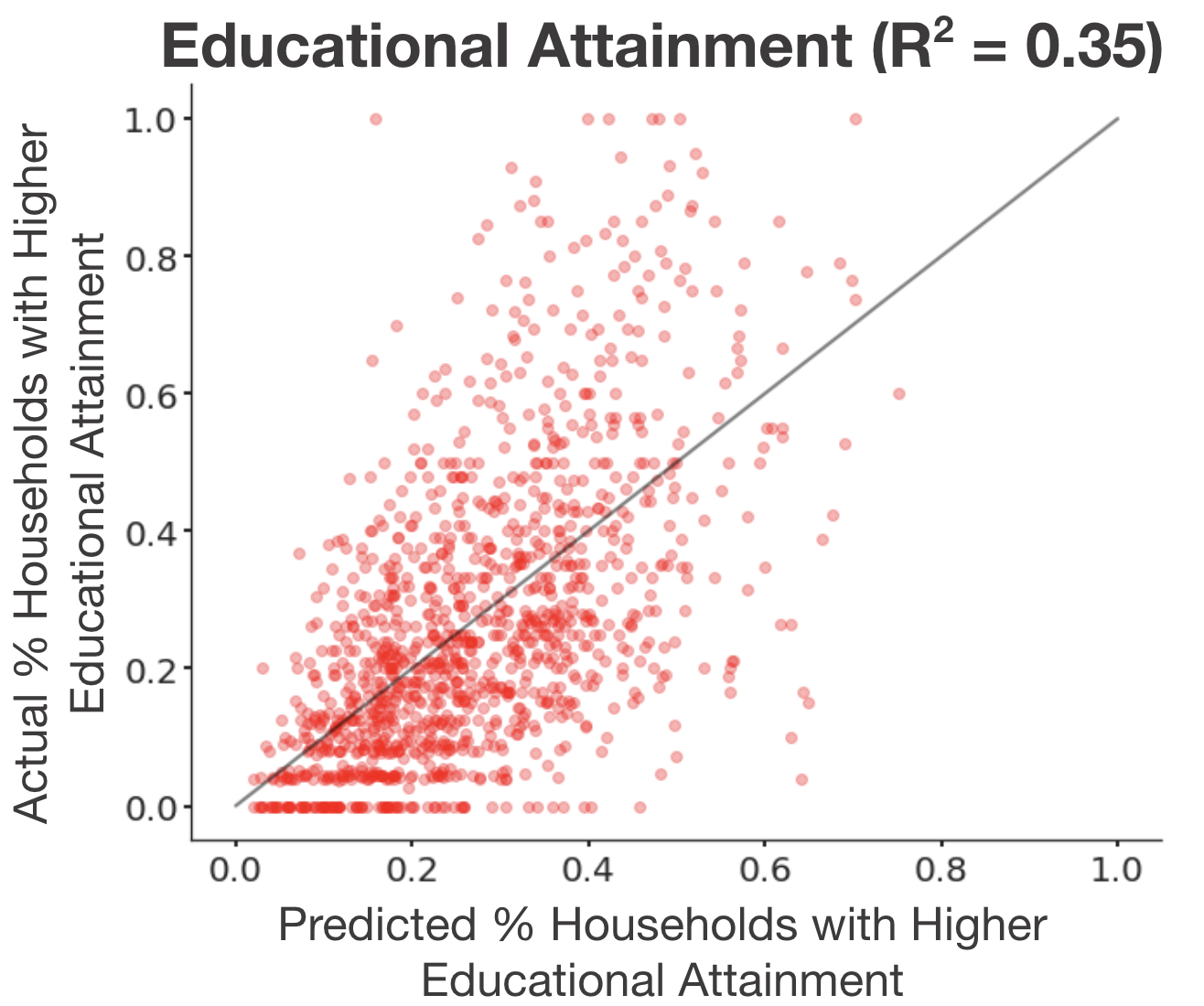}
  \end{subfigure}
  \caption{Actual and predicted 2017 Philippines DHS development indicators. }
\label{fig:results}
\end{figure*}

\end{appendices}

\end{document}